%Paper: hep-th/9209057
%From: "A. Jevicki" <MARYANNR@brownvm.brown.edu>
%Date: Wed, 16 Sep 92 11:34:20 EDT

\input phyzzx
\normalspace

%%%%definitions%%%
\def\delx{\partial_x}

\def\half{{1 \over 2}}
\def\a{\alpha}
\def\at{{\tilde\alpha}}
\def\ap{{\alpha_+}}
\def\am{{\alpha_-}}
\def\apm{{\alpha_{\pm}}}
\def\amp{{\alpha_{\mp}}}
\def\atp{{\tilde\alpha_+}}

\def\atpm{{\at_{\pm}}}
\def\atmp{{\at_{\mp}}}
\def\dpx{{\int {{dx} \over {2\,\pi}}\,}}

\def\dpt{{\int {{d\tau} \over {2\,\pi}}\,}}
\def\da{\int_{\am}^
{\ap} d\alpha}

\def\po{{\pi\phi_0}}

\def\mdot{{\dot M}}
\def\Tp{{T^
{(+)}_{ik,+}}}
\def\Tm{{T^
{(+)}_{ik,-}}}
\def\adag{{a_k^
{\dagger}}}

\FRONTPAGE
\line{\hfill BROWN-HET-874}
\line{\hfill CNLS-92-10}
\line{\hfill September 1992}
\bigskip
\title{{\bf
    SCATTERING STATES AND SYMMETRIES
    IN THE MATRIX MODEL AND TWO DIMENSIONAL STRING THEORY
     }}
\bigskip
\centerline{Antal Jevicki}
\centerline{\it Physics Department, Brown University,
Providence, RI 02912,
USA}
\centerline{Jo\~ao P. Rodrigues}
\centerline{\it Physics Department and Center for Nonlinear Studies,}
\centerline{\it University of the Witwatersrand, Wits 2050, RSA}
\centerline{Andr\'e J. van Tonder}
\centerline{\it Physics Department, Brown University, Providence,
RI 02912,
USA}
\centerline{\it and}
\centerline{\it Physics Department and Center for Nonlinear Studies,}
\centerline{\it University of the Witwatersrand, Wits 2050, RSA}
\bigskip
\abstract
{We study the correspondence between the linear matrix
model and the interacting nonlinear string theory.
Starting from the
simple matrix harmonic
oscillator states, we derive in a direct way
scattering amplitudes of 2-dimensional strings,
exhibiting the nonlinear equation generating
arbitrary N-point tree
amplitudes.
An even closer connection between the matrix model and the conformal
string theory is seen in studies of the symmetry algebra of the
system.
}

\vfill
\endpage

\chapter{Introduction}

\noindent
Matrix models
provide not only a novel formulation of low dimensional
string
theory but one which is integrable and exactly
solvable.
They lead to exact string equations in $D<1$ and a wealth
of results for the free energy and correlation functions [1].
The largest model understood so far is the two dimensional
string theory,
described by a
simple dynamics of a (matrix) harmonic oscillator [2-8].

While the numerical results follow straightforwardly,
the physical picture encoded
in the matrix model is however not seen directly. It is exhibited once
appropriate physical observables (collective fields)
[3] are identified.
For the tachyon one has the bosonic collective field defining
perturbative states. While the matrix model
is linear, the collective field
exhibits a nonlinear interaction which leads
to nontrivial physical scattering processes
[4].
A fermi liquid
description can be used to give a semiclassical
picture of the scattering
[5].
The field theory is integrable: it exhibits an infinite
sequence of conserved charges and an even larger symmetry of
$W_{\infty}$
generators
[6].

String theory is however most naturally described in terms of the
world sheet string coordinates and
associated conformal vertex operators [9].
These indeed exhibit similar symmetries [10] and can be seen to
give the
same correlation functions.  Except for the coincidence of various
results a closer connection between the matrix model description and
the string language is still lacking.

It is the purpose of this paper to address this problem and give
a more direct relationship between linear states of the matrix model
and nonlinear scattering states of string theory.
One has the matrix harmonic
oscillator

$$
           L = \half {\rm Tr} \bigl( \mdot^2 + M(t)^
2 \bigr),
           \eqno\eq
$$

\noindent
with

$$     \eqalign {
               A_{\pm} &\equiv P \pm M = \dot M \pm M, \cr
               A_{\pm}(t) &= A_{\pm}(0)\, e^
{\pm t}  \cr
}
$$

\noindent
being standard creation-annihilation operators. In terms of
these one easily
 writes down the eigenstates of the hamiltonian

$$
                 H = \half\, {\rm Tr\,} \bigl( (P+M)(P-M) \bigr) =
                     \half\, {\rm Tr\,} A_{+} A_{-}.
$$

\noindent
For example, the one-parameter set

$$
        A_{n}^{\pm} = {\rm Tr\,} (P\pm M)^
n
$$

\noindent
gives imaginary eigenvalues with energies
$\epsilon_n = \mp i\,n$.
Real energy
states are obtained by analytic continuation $n=ik$:

$$
        B_{k}^{(\pm )} = {\rm Tr\,} (P\pm M)^
{ik}.   \eqn\continuedstates
$$

The question is then how this simple set of exact matrix model
states translates into nontrivial string scattering states.
Continuing on the constructions begun in [6], we
shall explain a correspondence in section 2 and
describe a simple
derivation of general string scattering amplitudes using the
integrable states.
As such we exhibit how the
nonlinear string dynamics follows from the linear and
integrable matrix dynamics.

In section 3 we discuss the symmetry algebra of the theory.
We demonstrate there a close connection between the matrix
$W_\infty$ generators and those of the conformal string theory.
In particular we shall see that the collective (tachyon) field
representation of these operators is nothing but the representation
defined in the conformal approach by Klebanov in [11].

\chapter{From States to Scattering}

\noindent
Strings in two dimensions are described by the coordinates
$X^\mu \equiv (X, \phi)$, where $X$ is (usually) taken
as spacelike and $\phi$ is the nontrivial Liouville coordinate [9].
One has translation invariance in the $X$ direction (this is the
time coordinate of the matrix model, \ie $X = it$) and only
asymptotic translation invariance in $\phi$ due to an
exponential wall $\mu\, e^{-\sqrt{2}\phi}$.  The vertex operators
of the lowest string modes (massless tachyons) are

$$
\eqalign{
  V_{\pm} &= e^{ipX + \beta_{\pm} \phi},   \cr
  \beta_{\pm} &= -\sqrt 2 \pm |k|.
}  \eqno\eq
$$

\noindent
Only the $+$ branch describes physical scattering states.  The
$-$ operators grow at $\phi\to -\infty$ and are termed \lq\lq wrongly
dressed". For scattering one has left movers (as initial states)
and right movers (as final states) respectively denoted by

$$
  T^{(\pm)}_k = e^{ikX + (-\sqrt 2 + |k|)\phi},
  \eqno\eq
$$

\noindent
where $\pm = {\rm sign\,} k$.

States of the matrix model can be seen to be in close correspondence.
In particular, of \continuedstates\ half of the states have a
scattering interpretation as

$$ \eqalign {
       B_{-k}^{(-)} |0\rangle &= {\rm Tr\,} (P-M)^
{-ik} = |k\,; {\rm in}\rangle, \cr
       B_{k}^{(+)} |0\rangle  &= {\rm Tr\,} (P+M)^
{ik} = |k\,; {\rm out}\rangle, \cr
} \eqno\eq
$$

\noindent
This physical interpretation will arise once the spatial
(Liouville) coordinate is identified.
This was understood to be related to the eigenvalue index
of the matrix variable.  The
physical world is the positive real axis with a barrier at the
origin, and so one only considers an in state that is left moving and
an out state that is right moving.

The identification of physical states and of the extra Liouville
momentum is seen in a
transition to the collective field theory language [3].
This transition can be summarized [3-6] by the
following set of replacement rules:

$$ \eqalign {
                       M &\to x,      \cr
                       P &\to \a (x,t),   \cr
                     {\rm Tr} &\to \dpx \da.  \cr
}  \eqno\eq
$$

\noindent
The matrix hamiltonian then becomes

$$
    H  = {1 \over 6} \dpx \bigl( \ap^3 - \am^
3 \bigr)
       - \half \dpx x^
2\, \bigl( \ap -  \am \bigr),
    \eqno\eq
$$

\noindent
describing a scalar field $\phi(x,t)$ and its conjugate $\Pi (x,t)$,
with
$
       \apm = \delx \Pi\,\pm\,\pi\phi.
$

The collective representation exhibits in addition to the time $t$ a
spatial dimension $x$. One has a classical background field
$  \po = \sqrt{x^
2 - 2\mu} $, which induces a reparametrization of the
new spatial coordinate to
$
            \tau = \int{dx \over \po(x)},
$
or

$$  \eqalign {
          \quad x(\tau)  &= \sqrt{2\mu}\, \cosh(\tau),\cr
          \po (\tau)     &= \sqrt{2\mu}\, \sinh(\tau).\cr
}                 \eqn\Trajectory
$$

\noindent
Asymptotic translations in $\tau$ are scale transformations of
$x$ since

$$
                 x(\tau) \sim \sqrt{{\mu \over 2}}\,e^
{\tau}.
    \eqn\AsTrajectory
$$

\noindent
Indeed, in
addition to time translation the collective
Lagrangian transforms covariantly under
scale transformations

$$  \eqalign {
                    x &\to \lambda\,x, \cr
    \a (x,t) &\to {1 \over \lambda}\, \a (\lambda\,x,t), \cr
                    H &\to {1 \over \lambda
^4}\, H. \cr
} \eqno\eq
$$

\noindent
This symmetry is the origin of a second (spatial momentum)
quantum number $p_{\tau}$.

In linearized approximation with

$$ \eqalign {
   &\phi(x) = \phi_0(x) + \delx\,\psi(x), \qquad
              p(x) = - \delx\Pi(x), \cr
   &\ \psi(\tau) = \psi(x), \qquad \qquad p(\tau) = \po\, p(x), \cr
            & \apm(x) = \pm\,\po + {1 \over \po}\, \atpm(\tau). \cr
} \eqn\ChangeVarE
$$

\noindent
one has right-left moving massless modes (tachyons)

$$ \eqalign {
 \atpm(\tau,t)\,&= f(t\,\mp \,\tau) =
   \pm\int_{-\infty}^{\infty}dk\,\alpha^{\pm}_k\, e^
{-ik(t\mp \tau)},\cr
}    \eqn\LeftRight
$$

\noindent
satisfying

$$
         (\partial_t \pm \partial_{\tau}) \atpm  = 0.
         \eqno\eq
$$

\noindent
with the energy momentum values

$$
    \alpha^
{\pm}_{-k} : \quad p_0 = k, \quad p_{\tau} = \pm k.
    \eqno\eq
$$

The exact states of the matrix model are directly translated into the
field theoretic representation. We have as exact tachyon eigenstates

$$
            T^{(\pm)}_n = \dpx \int d\a\, (\a \pm x)^
{n} =
            \dpx { (\a \pm x)^
{n+1} \over n+1},
            \eqno\eq
$$

\noindent
introduced by Avan and one of the authors in [6].
Using the Poisson brackets $\{\a (x),\a (y)\} = 2\pi\, \delta '(x-y)$
one easily shows

$$
             \{H,T^{(\pm)}_n\} = \pm n\,T^
{(\pm)}_n
             \eqno\eq
$$

\noindent
and one has eigenstates with $ip_0 = \pm n$. Defining

$$
      p_{\tau} = {\it scale\  dimension\/} - 4, \eqno\eq
$$

\noindent
one has

$$
                        p_{\tau} = - 2 + n.
$$

\noindent
These states stand in comparison with
the vertex operators of conformal field
theory

$$
\eqalign{
              T_p^{(\pm)} &\equiv e^
{ i\,p\,X + (-\sqrt 2 + |p|)\varphi}  \cr
  &\leftrightarrow \dpx {(\alpha \pm x)^{n+1} \over n+1}  \cr
  &\leftrightarrow {\rm Tr \,} (P\pm M)^n.  \cr
}  \eqno\eq
$$

\noindent
The tachyon vertex operators with opposite (Liouville)
dressing correspond to singular
operators in the matrix model

$$
\eqalign{
  e^
{ i\,p\,X + (-2 - |p|)\,\varphi} &\leftrightarrow
  \dpx { (\a \pm x)^
{1-n} \over 1-n} \cr
   &\leftrightarrow   {\rm Tr\,} (P\pm M)^{-n}. \cr
} \eqno\eq
$$

We have now described a one to one correspondence between the matrix
model states and string states.  Scattering amplitudes can be
derived immediately once this correspondence is understood.

We note that the collective field theory seemingly introduces a
degeneracy. For each state of the matrix model one can define
two states
in collective field theory since we can replace $P \to \apm
(x,t)$. Each
of the separate fields $\ap$ or $\am$
can be used to define states with the
above quantum numbers. In particular

$$
    \dpx { (\ap \pm x)^
{1 \pm ik} \over {1 \pm ik} }
$$

\noindent
and

$$
    \dpx { (\am \pm x)^
{1 \pm ik} \over {1 \pm ik} }
$$

\noindent
both have the same quantum numbers

$$
           p_0 = k, \qquad p_{\tau} = -2 \pm ik.
$$

\noindent
These have to be identified, up to a phase factor.
It can be
shown (below) that boundary conditions fix the phase factor
to be $-1$. So one has

$$
    \dpx { (\amp \pm x)^
{1 \mp ik} \over {1 \mp ik} } =
    - \dpx { (\apm \pm x)^
{1 \mp ik} \over {1 \mp ik} },
    \eqn\Identification
$$

\noindent
implying a nonlinear relation between left and right movers.
This equation, which
follows from simple kinematical reasoning, determines the complete
tree level scattering amplitude. Expanding

$$
   \apm(x) = \pm\,x \mp {1 \over x}
             \bigl(\mu \mp \hat\alpha_{\pm}(\tau) \bigr) +
{1 \over x^
2}
              {\rm \ terms},
              \eqn\asymptotic
$$

\noindent
we shall find the relation

$$
\eqalign {
   \int_{-\infty}^\infty d\tau\, e^{\pm ik\tau}\,
        {{\hat\alpha}_{\pm}\over\mu}
       &=
       - \int_{-\infty}^{\infty} {d\tau \over ik\pm 1}\, e^
{\mp ik\tau}
       \Biggl[ \Bigl[ 1 + {\hat\alpha_{\mp} \over \mu} \Bigr]^
{ik\pm 1} - 1
             \Biggr].\cr
} \eqn\Moore
$$

\noindent
This functional equation relating left and right moving waves
of the collective
field was shown to represent a solution to the
scattering problem in
[8].  Here we exhibited how this nonlinear scattering equation
emerges directly from the
exact oscillator states.
The fact that the left and the right hand side of the equation
are interpreted as eigenstates of collective field theory
implies also the following:  a complete quantization procedure was
given [4] for the field theory Hamiltonian, involving normal
ordering and the subtraction of counterterms.  The same procedure
can be applied to the states and will lead to a fully quantum
version of the scattering equation.

The main ingredient in obtaining the scattering equation are
the proper boundary conditions.
Let us now elaborate on this question.
The issue of boundary conditions is of paramount
importance in a correct treatment of
the spectrum within the collective
approach. In QCD-like unitary matrix models,
it is well known that as the system moves from
a strong coupling regime to a weak coupling regime where the classical
density of states $\phi_0$ has only finite support, Dirichlet boundary
conditions must be imposed on the shifted field $\psi(\tau)$. This is
essentially due to the fact that $\phi_0(\tau=0)=\phi_0(\tau=L
\to\infty)=0$,
and in this way
the time independence of the original constraint condition $\int\,dx\,
 \phi=N$
is preserved
[3]. For $c=1$ strings,
this \lq\lq constraint" equation determines the value of the
cosmological constant. Therefore, apart from problems of consistency,
a choice other than Dirichlet boundary conditions would result in a
time
dependent cosmological constant.
Notice that this implies that in a density
variable description of \lq\lq wall"
scattering, the \lq\lq wall" at $\tau=0$ is
rigid.
A creation-annihilation basis that automatically enforces Dirichlet
boundary conditions on the scalar field $\psi$ is defined by the
expansion

$$ \eqalign{
  &\atpm(\tau)\,= \pm\,\int_{-\infty}^
{\infty}{dk \over \sqrt{|k|}}\,
  e^
{\pm\,i\,k\tau} a_k , \quad a_{-k}\equiv \adag,\cr
  &
[a_k,\adag] = \delta (k-k').\cr
}   \eqn\StandExp
$$

\noindent
We could equally well have chosen the \lq\lq left-right" basis
\LeftRight.
Once one expresses a scalar theory with fields satisfying boundary
conditions in a left-right basis, there is a standard problem, also
present in the critical open string: the functions $e^
{ik\tau}$ are not
orthogonal over the half line, and therefore the computation of
Fourier
coefficients require some modification. To this standard problem
there
is a standard solution
[12]: one notices that
the definitions of all the fields in \LeftRight\ naturally extend to
negative values of $\tau$. Therefore we extend the definition of
the fields from $ 0 \le \tau < \infty$ to $-\infty < \tau < \infty$
by requiring

$$   \eqalign {
            \psi(-\tau) &= - \,\psi(\tau),\cr
            \atpm(-\tau) &= - \,\atmp(\tau).\cr
}      \eqn\Involut
$$

\noindent
In other words, the fields of interest to us are the fields defined
on the
full line which are odd (in coordinate free form) under the involution
$\tau \to -\tau$. This point of view has been extensively used in works
relating critical open string amplitudes to those of the closed string
[13].
One can then compute Fourier coefficients of $\atp$, say:

$$
    \int_0^{\infty} {d\tau \over 2\pi}\, e^
{\mp\,ik\tau}\, \atpm(\tau)-
    \int_0^{\infty} {d\tau \over 2\pi}\, e^
{\pm\,ik\tau}\, \atmp(\tau) =
    \pm \alpha_k.
    \eqn\Extend
$$

\noindent
We can now reformulate the problem as follows:
suppose we introduce the arbitrary left-right expansion \LeftRight.
Equation \Involut\ is then equivalent to

$$
            \alpha^-_k = - \alpha^
+_k.     \eqn\BigEq
$$

\noindent
Physically, this simply means that in order to preserve the boundary
conditions of the system,
if a right mover is created then a left mover must also be created
with
amplitude minus one, and similarly for annihilation operators.
This means that the Dirichlet boundary conditions cause the
 left and right
movers to combine into standing waves, which are perturbative tachyon
states in the matrix model.

Now, in terms of the matrix variables  \StandExp\
described above, this condition
is immediately built into the expansion of the fields. However, for
asymptotic incoming and outgoing states,
which are naturally defined on the full line,
the analogue of condition \BigEq, imposed on the
the exact states of the system, leads to the nonlinear scattering
matrix.

We now concentrate on $T^
{(+)}_{ik}$ and introduce the following notation
to represent the two degenerate states described previously

$$ \eqalign {
      T^{(+)}_{ik} &= \dpx \da\,(\a+x)^
{ik}
                   \equiv T^{+}_{ik\,(+)} - T^
{+}_{ik\,(-)} \cr
     &= \dpx { (\amp + x)^
{ik + 1} \over {ik + 1} } -
        \dpx { (\apm + x)^
{ik + 1} \over {ik + 1} }.\cr
}   \eqn\ExactDefX
$$

\noindent
The equation relating left and right moving fields reads

$$  \eqalign {
     \Tp &= \half \int_{-\infty}^
{\infty} {d\tau \over 2\pi}\,
           {\po \over ik+1} \Biggl[
           \Bigl( (\po+x) + {\ap(\tau) \over \po} \Bigr)^
{ik+1}
             - x^
{ik+1} \Biggr] \cr
         &= - \half \int_{-\infty}^
{\infty} {d\tau \over 2\pi}\,
           {\po \over ik+1} \Biggl[
           \Bigl( (-\po+x) + {\am(\tau) \over \po} \Bigr)^
{ik+1}
             - x^
{ik+1} \Biggr] \cr
         &= - \Tm.\cr
}   \eqn\IdentificationB
$$

\noindent
The range of integration has been extended as described above equation
\Extend. This is a restatement of equation \Identification.
Since the c-number contributions $C_{\pm}^
{(+)}$ to the above operators are
are the same, we rewrite this condition as

$$
       \Tp - C_{+}^{(+)} = - \bigl( \Tm - C_{-}^
{(+)} \bigr).
       \eqn\TheCondition
$$

\noindent
This equality is a necessary consequence of Dirichlet boundary
conditions.
To linear order, it is straightforward to show that
equation \TheCondition\ is equivalent to equation \BigEq .

As $x \gg \sqrt{2\mu}$ we wish to express this condition in terms
of the asymptotic fields $\hat \alpha_\pm$ defined in equation
\asymptotic, using the asymptotic behaviour
\AsTrajectory.
We remind ourselves that
$\hat\alpha_{-}(t+\tau)$ is an incoming left-moving wave
and $\hat\alpha_{+}(t-\tau)$ is the outgoing, right-moving wall
scattered
wave.
In
the asymptotic description the fact
that, from the collective field theory point of view, the \lq\lq wall"
at
$\tau=0$ is rigid, is not immediately built into the definition of
the fields.
This condition has to be imposed on the exact states
of the system, i.e., equation
\TheCondition\  must be satisfied.  Expressing the exact states in
terms of the variables \asymptotic\ we get

$$  \eqalign {
    \Tp - C_{+}^
{(+)}
        &= {1\over ik+1}\,\dpx {(\ap+x)^{ik+1} - (\po+x)^
{ik+1}}\cr
        &= { \sqrt{2\mu}^{\,ik+1}\over ik+1} \int_{-\infty}^
{\infty}
{d\tau \over 8\pi}\,
           {e^{ik\tau}}\, e^
{2\tau} \cr
   &\ \Biggl[
                \sum_{j=0}^
\infty
                {\Gamma(ik+2) \over  \Gamma(ik + 2 - j)\,j!}
                \,(-)^j\,e^
{-2j\tau}
      \left\{\left(1-{\hat\alpha_+\over\mu}\right)^
j-1\right\}
             \Biggr],  \cr
}  \eqno\eq
$$
$$
\eqalign{
   - \bigl( \Tm - C_{-}^
{(+)} \bigr)
        &= - {1\over ik+1}\dpx (\am+x)^{ik+1} - (-\po+x)^
{ik+1} \cr
        &= - {\sqrt{2\mu}^{\,ik+1}\over ik+1} \int_{-\infty}^
{\infty}
              {d\tau \over 8\pi} \,
           e^{-ik\tau} \sum_{p=1}^
\infty
                {\Gamma(ik+2) \over  \Gamma(ik + 2 - p)\,p!}
                \left( {\hat\alpha_- \over \mu} \right)^
p.  \cr
} \eqno\eq
$$

\noindent
Equating these expressions as required by the condition
\TheCondition, and applying partial integrations and a Fourier
transform, we obtain

$$  \eqalign {
        &
                \sum_{j=0}^\infty e^
{-2(j-1)\tau}\,
                {\Gamma(-\partial+2) \over  \Gamma(-\partial + 2 -
j)\,j!}
                \,(-)^j\,e^
{-2j\tau}
  \left\{\left(1-{\hat\alpha_+(\tau)\over\mu}\right)^
j-1\right\} \cr
  &\ =
        -
           \sum_{p=1}^
\infty
                {\Gamma(\partial+2) \over  \Gamma(\partial + 2 -
p)\,p!}
                \left( {\hat\alpha_-(-\tau) \over \mu} \right)^
p.  \cr
} \eqno\eq
$$

\noindent
We can now extract the asymptotic limit by letting $\tau\to\infty$.
We find that on the left hand side only the $j=1$ term contributes
(lower order terms would correspond, on the right hand side, to
terms dropped in the asymptotic definition \AsTrajectory).
As $\tau\to \infty$

$$  \eqalign {
        &           (-\partial+1)
            {\hat\alpha_+(\tau)\over\mu}
\cr
  &\ =
        -
           \sum_{p=1}^
\infty
                {\Gamma(-\partial+2) \over  \Gamma(-\partial + 2 -
p)\,p!}
                \left( {\bar\alpha_-(\tau) \over \mu} \right)^
p,  \cr
} \eqno\eq
$$

\noindent
where $\bar\alpha_-(\tau) \equiv \hat\alpha_-(-\tau)$.  It follows
that $$
\eqalign {
        &            {\alpha_+(\tau)}
   =     -
           \sum_{p=1}^
\infty
                {\Gamma(-\partial+1) \over  \Gamma(-\partial + 2 -
p)\,p!} \, \left({1\over\mu}\right)^
{p-1}\,
                \bar\alpha_-(\tau)^
p.  \cr
} \eqno\eq
$$

This relation expressing left moving fields in terms of right moving
ones is the result \Moore\ for the scattering problem [8].

\chapter{Symmetries}

\noindent
The spacetime field theory given by the collective field exhibits a
large ($W_\infty$) spacetime symmetry of 2-dimensional string theory
[6].
The generators of this symmetry can be directly found or simply
induced from the matrix model.  There one has the invariant
operators

$$
  {\rm Tr\,} (P^r M^s),       \eqno\eq
$$

\noindent
which are closed under commutation.  The field theory operators
read

$$
  H_m^n =      \dpx {\alpha_+^{m-n}\over m-n}\,x^{m-1}  \eqno\eq
$$

\noindent
and can be shown to satisfy the $w_\infty$ algebra

$$
  [H^{n_1}_{m_1}, H^{n_2}_{m_2}]
   = i\, [(m_2-1)\,n_1 - (m_1-1)\,n_2]\, H^{n_1+n_2}_{m_1+m_2-2}.
\eqno\eq
$$

\noindent
Of particular relevance to us are the spectrum generating operators

$$
  O_{JM} \equiv {\rm Tr\,} (P+M)^{J-M}(P-M)^{J+M},   \eqno\eq
$$

\noindent
which become, in the collective field theory representation

$$
  O_{JM} =     \dpx\int_{\alpha_-}^{\alpha_+}
   d\alpha\, (\alpha+x)^{J+M+1}\,(\alpha-x)^{J-M+1}.
   \eqn \specgen
$$

\noindent
One sees that these are linear combinations of the basic
$w_\infty$ operators

$$
  O_{JM} = H_1^{-2J-2} + 2M\, H_2^{-2j} + (2M^2-J-1)\, H_3^{2J+2}
   + \dots.
\eqno\eq
$$

\noindent
and it follows that the spectrum generating algebra is precisely a
$w_\infty$, \ie,

$$
  [\,O_{J_1,M_1}, O_{J_2,M_2}\,]
   = i\, [(M_2-1)\,J_1 - (M_1-1)\,J_2]\, O_{J_1+J_2,M_1+M_2}.
  \eqn\algebra
$$

\noindent
This can also be shown directly from \specgen\ by doing partial
integrations [6].

There is a close connection between these $w_\infty$ operators
and the operators describing exact tachyon states of the
field theory.  Recall
the one parameter family of operators

$$
  T^{(\pm)}_n =     \dpx\int_{\alpha_-}^{\alpha_+}
   d\alpha\, (\alpha\pm x)^n.
\eqno\eq
$$

\noindent
Their commutators give the generators of the $w_\infty$ algebra,
\ie, one can show that

$$
  O_{JM} = {1\over 2i\,(J-M+2)\,(J+M+2)}\, [\,T^+_{J+M+2},
    T^-_{J-M+2}\,].
\eqno\eq
$$

\noindent
The symmetry generators were also written down in the
conformal field theory approach [10].
There is a close parallel with all of
the matrix model relationships
and the commutators are
simply replaced by operator products.  The above implies for example
that the
$w_\infty$ generators are obtained as operator products of basic
tachyon vertex operators.
A closer correspondence is seen by comparing the above field
theory forms with  the representations deduced for the action of the
symmetry generators on the tachyon module [11].

\section {Fourier Expansion}

\noindent
We now
consider the spectrum generating operators $O_{JM}$ of \specgen\
in more detail.
We shall see that the correspondence with the conformal field theory
results
of [11] will then follow.   To expand in terms of
creation-annihilation
operators we make the substitutions

$$
\eqalign{
  \alpha_+ &= x+ \bar\alpha_+,  \cr
  \alpha_- &= -x + \bar\alpha_- \cr
} \eqn\defA
$$

\noindent
in the spectrum generating operators \specgen.
Applying partial integration to \specgen\
and inserting the limits \defA, one finds

$$
\eqalign{
  O_{JM} =      \dpx
         \sum_{k=0}^{J+M+1} (-)^k\,
          & {(J-M+1)!\,(J+M+1)! \over (J-M+2+k)! \,(J+M+1-k)!}
            \times\cr
         \times\bigl\{
           & {\bar\alpha_+}^{J-M+2+k}\,
           (\bar\alpha_+ + 2x)^{J+M+1-k}   \cr
          -
         &
           (\bar\alpha_- - 2x)^{J-M+1+k}\,
           {\bar\alpha_-}^{J+M+2-k}
           \bigr\}. \cr
}  \eqn  \partialA
$$

\noindent
The leading term in $\bar\alpha_+$ is of order
${\bar\alpha_+}^{J-M+2}$.  The leading term in $\bar\alpha_-$
seems to be linear in $\bar\alpha_-$.  However, this is not true, as
careful consideration shows that there are two terms linear
in $\bar\alpha_-$ which cancel.  One might expect that in general
something
similar happens also for higher order terms in $\bar\alpha_-$.
This indeed turns out to be the case.  The easiest way to see this,
is to do the partial integration of \specgen\ in the other
\lq\lq direction".  One finds

$$
\eqalign{
  O_{JM} =      \dpx
          \sum_{k=0}^{J-M+1} (-)^k\,
           &{(J+M+1)!\,(J-M+1)! \over (J+M+2+k)!\, (J-M+1-k)!}
           \times\cr
         \times\bigl\{
          &{\bar\alpha_+}^{J-M+1-k}\,
           (\bar\alpha_+ + 2x)^{J+M+2+k}   \cr
          -
          &(\bar\alpha_- - 2x)^{J-M+1-k}\,
           {\bar\alpha_-}^{J+M+2+k}
           \bigr\}. \cr
}  \eqn  \partialB
$$

\noindent
The terms in $\bar\alpha_+$ and $\bar\alpha_-$ in \partialA\
and \partialB\ must separately be equal, up to c-number terms
of the form $\dpx x^{2J+3}$.  Substituting the
change of variables \AsTrajectory, this becomes
$\sim \int_{-\infty}^\infty {d\tau\over 2\pi}\, e^{(2J+3)\tau}$,
which can in general be argued to vanish after an analytic
continuation $\tau\to i\tau$ (see below).  It therefore follows
that we can write the expansion

$$
\eqalign{
  O_{JM} =
      &\dpx
         \sum_{k=0}^{J+M+1} (-)^k\,
           {(J-M+1)!\,(J+M+1)! \over (J-M+2+k)!\, (J+M+1-k)!}\,
           {\bar\alpha_+}^{J-M+2+k} \times  \cr
     & \qquad\qquad\qquad\qquad \times
           (\bar\alpha_+ + 2x)^{J+M+1-k}   \cr
           &- \dpx
          \sum_{k=0}^{J-M+1} (-)^k\,
           {(J+M+1)!\,(J-M+1)! \over (J+M+2+k)!\, (J-M+1-k)!}\,
           {\bar\alpha_-}^{J+M+2+k} \times  \cr
     & \qquad\qquad\qquad\qquad \times
           (\bar\alpha_- - 2x)^{J-M+1-k}.  \cr
}  \eqn  \fullCharge
$$

\noindent
Thus to lowest order in the fields, one finds

$$
\eqalign{
  O_{JM} &= {1\over J-M+2}     \dpx (2x)^{J+M+1} \,
            {\bar\alpha_+}^{J-M+2}  \cr
       &\ - {1\over J+M+2}     \dpx (-2x)^{J+M+1} \,
            {\bar\alpha_-}^{J+M+2}. \cr
}  \eqno\eq
$$

Now, applying the change of variables \AsTrajectory, \ie,

$$
\eqalign{
  x &= \sqrt {\mu\over 2}\, e^\tau,      \cr
  \bar\alpha_\pm &\to {d\tau\over dx} \bar\alpha_\pm,  \cr
}  \eqno\eq
$$

\noindent
one finds that the leading order expression for the charges is
given by

$$
\eqalign{
  O_{JM} &= {{2}^{J+1}{\mu}^{M}
  \over J-M+2}    \, \dpt
  e^{2M\tau}\,
            {\bar\alpha_+}^{J-M+2}  \cr
       &\ - (-)^{J-M+1}\,
       {{2}^{J+1}\, {\mu}^{-M}
       \over J+M+2}\,
           \dpt e^{-2M\tau}\,
            {\bar\alpha_-}^{J+M+2}. \cr
}  \eqno\eq
$$

\noindent
Expanding in right and left moving modes

$$
\eqalign{
  \bar\alpha_+ &= \int_{-\infty}^\infty
       dk\, \bar\alpha(k)\, e^{-ik(t-\tau)},  \cr
  \bar\alpha_- &= \int_{-\infty}^\infty
       dk\, \bar\beta (k)\, e^{-ik(t+\tau)}  \cr
}    \eqno\eq
$$

\noindent
and applying the rotation $\tau\to i\tau$, $k\to -ik$,
we find that in terms of the analytically continued oscillators

$$
\eqalign{
   \alpha(k) &\equiv \bar\alpha(-ik),  \cr
   \beta(k) &\equiv \bar\beta(-ik)   \cr
} \eqn\defB
$$

\noindent
the charges have the form

$$
\eqalign{
  O_{JM} = &{{2}^{J+1}{\mu}^{M}
  \over J-M+2}\, i   \int dk_1 \dots dk_{J-M+2}\, \times \cr
    &\qquad \times\alpha (k_1) \dots \alpha (k_{J-M+2}) \,
      \delta\left(\sum k_i + 2M\right)  \cr
       &\ - (-)^{J-M+1}\,
       {{2}^{J+1} {\mu}^{-M}
       \over J+M+2}\,i  \int dp_1 \dots dp_{J+M+2}\,\times  \cr
    &\qquad \times\beta  (p_1) \dots \beta  (p_{J+M+2}) \,
      \delta\left(\sum p_i + 2M\right).  \cr
}  \eqn\fullrep
$$

We emphasize that this is the expression
for the charges to lowest order in the fields, which corresponds to the
leading order in $\mu$. The full expression \fullCharge\ has
corrections in $1/\mu$ that are higher order polynomials in the
fields.  In the remainder of the discussion we do not consider
these corrections.

Defining

$$
\eqalign{
  a(k) &\equiv \alpha(k), \qquad b(p) \equiv \beta(p),  \cr
  a^\dagger(k) &\equiv \alpha (-k)/k,
  \qquad b^\dagger(p) \equiv \beta(-p)/p  \cr
}  \eqno\eq
$$

\noindent
satisfying $[a(k), a^\dagger(k')] = \delta (k - k')$,
         $[b(p), b^\dagger(p')] = \delta (p-p')$,
we have the expressions of [11] (up to an inessential difference
in normalization),
plus additional contributions.  To see
these, note that in addition to the term

$$
\eqalign{
             {2}^{J+1}{\mu}^{M}\,
               i  & \int_0^\infty dk
               \int_0^\infty dk_1 \dots dk_{J-M+1}\times  \cr
  &\quad \times k\,a^\dagger(k)\,
    a (k_1) \dots a (k_{J-M+1}) \,
      \delta\left(\sum k_i - k + 2M\right)  \cr
}
\eqn \termA
$$

\noindent
found in [11], we in general also have terms of higher
order in the creation operators.  The next term would be, for
example

$$
\eqalign{
             {2}^{J+1}{\mu}^{M} (J-M  &  +1)\,
               i  \int_0^\infty dk\,dk'
               \int_0^\infty dk_1 \dots dk_{J-M  }\,\times  \cr
   &\times  kk'\,a^\dagger(k)\, a^\dagger(k')\,
    a (k_1) \dots a (k_{J-M }) \,
      \delta\left(\sum k_i - k - k'+ 2M\right).  \cr
}
\eqn\termB
$$

\noindent
If $M<0$, we also get an additional contribution of the form

$$
  \eqalign{
            {{2}^{J+1}{\mu}^{M}\over J-M+2}\,
               i
               &\int_0^\infty dk_1 \dots dk_{J-M+2}\,\times  \cr
    &\ \times a (k_1) \dots a (k_{J-M+2}) \,
      \delta\left(\sum k_i + 2M\right).  \cr
}
\eqn \termC
$$

These additional contributions have to be included in order
to obtain a representation of the algebra  \algebra.  The reason
for this is that terms of the type \termB, commuted
with terms of the type \termC, give additional contributions of
the type \termA, which are needed to again obtain a member of the
algebra on the right hand side.  This effect cannot be produced
by only using terms of the type \termA.  To see where the
representation \termA\ fails, one has to take careful account of
the regions of momentum integration.  For example, if one were
to use only terms of the type \termA\ one would find for the
commutator

$$
\eqalign{
 \left[ O_{MM}, O_{\half,-\half} \right]
       =  &\int_0^\infty dk_1 dk_2\,
          (-2k_1 - 2k_2 - 4M)\, (k_1+k_2+M-\half) \times \cr
    &\qquad\qquad\times a^\dagger (k_1+k_2+M-\half)\,a(k_1)\,a(k_2)
\cr
         &+\int_{0,k_1+k_2>\half}^\infty dk_1 dk_2\,
          (2k_1 + 2k_2 - 1)\, (k_1+k_2+M-\half) \times  \cr
  &\qquad\qquad\times a^\dagger (k_1+k_2+M-\half)\,a(k_1)\,a(k_2).
\cr
}
\eqno\eq
$$

\noindent
which would give $(-4M-1)\, O_{M+\half,M-\half}$, were it not for
the fact that the regions of integration do not match.  It is now
not difficult to see how to fix the representation \termA.  Simply
remove the restrictions on the ranges of integration, \ie, take
them to be $\int_{-\infty}^\infty dk$ instead of
$\int_{0}^\infty dk$.  This solves the problem on a formal level,
and imposing the reality conditions $a_{-n} = n a^\dagger_n
\equiv \alpha_{-n}$, we recover our full representation \fullrep.

One can now ask whether the Ward identities derived in [11] for
the tachyon scattering amplitudes will be affected by these
corrections.  As we will show in the next section, they will
not be affected.

\section {Ward Identities}

\noindent
One can now identify the spectrum generating operators
as we did for the tachyons by comparing quantum numbers as
in \Identification, or alternatively, by imposing Dirichlet
boundary conditions as in \IdentificationB.
One simply requires

$$
  O_{JM,+} = - O_{JM,-},   \eqno\eq
$$

\noindent
which implies that to leading order

$$
\eqalign{
  O_{JM} &= 2i\,{{2}^{J+1}{\mu}^{M}
  \over J-M+2}\,   \int dk_1 \dots dk_{J-M+2}\times  \cr
    &\qquad \times \alpha (k_1) \dots \alpha (k_{J-M+2}) \,
      \delta\left(\sum k_i + 2M\right)  \cr
       &= 2i\,(-)^{J-M+1}\,
       {{2}^{J+1} {\mu}^{-M}
       \over J+M+2}\,  \int dp_1 \dots dp_{J+M+2} \times \cr
    &\qquad \times\beta  (p_1) \dots \beta  (p_{J+M+2}) \,
      \delta\left(\sum p_i + 2M\right).  \cr
}  \eqn\specgenIdent
$$

\noindent
This identification
will, in practice, be very useful in
explicit calculations of
Ward identities, as will be seen below.

The \lq\lq bulk" scattering amplitudes only involve
fixed, discrete values of the outgoing momenta.  This can be
interpreted in our formalism as follows:  Imposing the above
identification of quantum numbers, one has

$$
\eqalign{
  T^{(-)}_{2M} \equiv O_{M-1,M}
    &= 2i\, (-)^{2M+1} \left( {2\over\mu}\right)^M \beta(2M)  \cr
    &= 2i\, (2\mu)^M \int dk_1\dots dk_{2M+1}\,
        \alpha(k_1) \dots \alpha (k_{2M+1})
         \,\delta (\sum k_i - 2M).   \cr
}  \eqn\Tident
$$

Thus, in terms of the oscillators $\alpha(k)$ and $\beta(p)$
defined in
\defA\ and \defB, we find an S-matrix that is different from the
one we previously calculated in terms of the \lq\lq asymptotic"
variables \AsTrajectory.  In particular, to leading order
an out state $\langle 0|\beta (2M)$ is $(2M+1)$-linear in
$\alpha$, so that a correlation function
$\langle\, \beta(p)\alpha(k_1)\dots\alpha(k_N)\,\rangle$ can only
be nonvanishing to this order if

$$
  p = N-1.   \eqn \sumrule
$$

\noindent
Except for an overall factor of $\half$, due to different normalization
of the momentum, this agrees with the \lq\lq sum rule" stated in [11].

Now, to see how the Ward identities can be derived in our formalism,
note that if one has an operator $O$ that annihilates the vacuum from
the left and the right, \ie,

$$
  \langle 0| O = 0 = O |0\rangle,   \eqno\eq
$$

\noindent
then, starting from the expectation value

$$
  \langle \,\beta(p)\, O \,\alpha(k_1) \dots \alpha(k_N)\,\rangle
   \eqn\expectation
$$

\noindent
one can write, commuting $O$ respectively to the left and to
the right

$$
  \langle\, [\beta(p),O] \,\alpha(k_1) \dots \alpha(k_N)\,\rangle
    =
  \langle\, \beta(p)\, [O, \alpha(k_1) \dots \alpha(k_N)]\,\rangle.
 \eqn  \wardId
$$

\noindent
This equation expresses the Ward identities, and for suitable
choices of the charges $O$, can be used to derive recursion relations
relating scattering amplitudes.

In general, however, our representation \fullrep\ of the
charges $O_{JM}$ have terms of the type $aa\dots a$ or $a^\dagger
a^\dagger\dots a^\dagger$, and therefore would fail to annihilate the
vacuum from either the left or the right.  Also, in addition to the
terms \termA, which were considered in the analysis of [11], one
might expect corrections to the Ward identities derived in [11]
due to our extra terms such as \termB\ and \termC.  However, counting
numbers of creation and annihilation operators, one sees that indeed
only terms of the type \termA\ contribute to the Ward identities.
For example, consider the Ward identity relating $N+1\to 1$
amplitudes to $N\to 1$ amplitudes.  The relevant charge is
$O_{\half,-\half} \sim aaa + a^\dagger a a + a^\dagger a^\dagger a$,
and we should take the momentum of the out state to be $p=N-1$.
The out state is then, from our previous discussion, of order

$$
\langle 0|\,\beta(p) \sim \langle 0|a^N,  \eqno\eq
$$

\noindent
so that one can write \expectation\ as

$$
  \eqalign{
  &\langle \,\beta(p)\, O_{\half, -\half}\,
  \alpha(k_1) \dots \alpha(k_{N+1})\,\rangle  \cr
  &\ \sim \langle a^N \,( aaa + a^\dagger aa + a^\dagger
       a^\dagger a )\, (a^\dagger)^{N+1}\rangle. \cr
}  \eqno \eq
$$

\noindent
It immediately follows that only the term linear in $a^\dagger$
contributes.  This argument generalizes to the identity for
expressing $N\to 1$ amplitudes directly in terms of $2\to1$
amplitudes, where the relevant charge is $Q_{N/2-1, -N/2-1}$.
The conclusion is therefore that for the purpose of deriving
these Ward identities, it
is sufficient to consider only the terms \termA\ linear in
the creation operators, as was done in [11].

Finally,
as an example, we calculate the Ward identity relating
$3\to 1$ amplitudes to $2\to 1$ amplitudes.  Using the
identification \specgenIdent, we have

$$
\eqalign{
  Q_{\half,-\half}
   &= {4\sqrt2\, i\over \sqrt\mu}
      \int_0^\infty dk_1 dk_2 dk_3\,\, k_1 \,a^\dagger (k_1)\,
         a(k_2)\, a(k_3)\, \delta(-k_1+k_2+k_3-1)  \cr
   &= {4\sqrt{2\mu}\, i}
      \int_0^\infty dp_1 dp_2 \,\, p_1 \,b^\dagger (p_1)\,
         b(p_2)\, \delta(-p_1+p_2-1).  \cr
}  \eqno\eq
$$

\noindent
up to terms that we have argued to be irrelevant.
Inserting this into the general formula \wardId, for $p=1$ and
$k_1+k_2+k_3 = 2$, we obtain the Ward identity

$$
  \langle\, b(2)\,a^\dagger(k_1)\,a^\dagger(k_2)\,a^\dagger(k_3)
  \,\rangle
  = {1\over\mu}\, (k_1+k_2-1)\,
  \langle\, b(1)\,a^\dagger(k_1+k_2-1)\, a^\dagger(k_3)\,\rangle
   + {\rm cyclic}.
$$

It is also possible to derive recursion relations in our formalism
using methods similar to those used in [11].  The argument
roughly goes as follows:  The operators

$$
  O_{NN} \equiv \int dx \int d\alpha\, (\alpha+x)^{2N+1} (\alpha-x)
$$

\noindent
have quantum numbers
$
  p_x = N = p_\tau,
$
while $T_k$ has $p_x = k$, $p_\tau = -1 + k$.
Adding these, it follows that the commutator $[\,O_{NN}, T_k\,]$
has quantum numbers $p_x = N+k$, $p_\tau = -1 + (n+k)$, and should
therefore be identified with $T_{k+N}$, \ie,

$$
  [\,O_{NN}, T_k\,] \sim T_{k+n}.  \eqno\eq
$$

\noindent
Similarly, the charge $O_{N+M,M}$ has quantum numbers
$p_x = M$, $p_\tau = N+M$.  Thus it follows that
$[\,O_{N+M,M}, T_{k_1}\ldots T_{k_{N+1}}\,]$ has quantum numbers
$p_x = M + \sum k_i$, $p_\tau = -1 + (m+\sum k_i)$, so that we have
to identify

$$
  [\,O_{N+M,M}, T_{k_1}\ldots T_{k_{N+1}}\,] \sim
  T_{M+\sum_{i=1}^{N+1}k_i}.  \eqno\eq
$$

In conlusion, we stress that the matrix model $w_\infty$ generators,
when expanded, give the conformal field theory expressions of [11]
and the associated bulk Ward identities.  But in addition they also
contain higher corrections in $1/\mu$, which could be
used to derive improved Ward identities which would
give the full amplitudes.

\vfill
\endpage

\centerline{\it REFERENCES}
\bigskip

\pointbegin
D. J. Gross and A. A. Migdal, {\it Phys. Rev. Lett.} {\bf 64} (1990)
127;
M. R. Douglas and S. Shenker, {\it Nucl. Phys.} {\bf B335} (1990) 635;
E. Br\'ezin and V. Kazakov, {\it Phys. Lett.} {\bf B236} (1990) 144.

\point
D. J. Gross and N. Miljkovi\'c, {\it Phys. Lett.} {\bf B238} (1990) 217;
E. Br\'ezin, V. A. Kazakov and A. B. Zamolodchikov, {\it Nucl.Phys.}
{\bf B338} (1990) 673;
P. Ginsparg and J. Zinn-Justin, {\it Phys. Lett.} {\bf B240} (1990) 333;
D. J. Gross and I. R. Klebanov, {\it Nucl. Phys.} {\bf B344} (1990) 475.

\point
S. R. Das and A. Jevicki, {\it Mod. Phys. Lett.} {\bf A5} (1990) 1639;
A. Jevicki and B. Sakita, {\it Nucl.Phys.} {\bf B165} (1980) 511.

\point
K. Demeterfi, A. Jevicki and J. P. Rodrigues, {\it Nucl.Phys.}
{\bf B362} (1991) 173; {\bf B365} (1991) 499; {\it Mod. Phys. Lett.}
{\bf A35} (1991) 3199.
\point
J. Polchinski, {\it Nucl.Phys.} {\bf B362} (1991) 25.

\point
J. Avan and A. Jevicki, {\it Phys. Lett.} {\bf B266} (1991) 35;
{\bf B272} (1992) 17;
D. Minic, J. Polchinski and Z. Yang, {\it Nucl. Phys.} {\bf B369}
(1992) 324;
M. Awada, S. J. Sin; UFIFT (Florida)-HEP 90-33 and 91-03;
G. Moore, N. Seiberg; {\it Int. J. Mod. Phys.} {\bf A7} (1992) 2601;
S. R. Das, A. Dhar, G. Mandal and S. Wadia;
   {\it Mod. Phys. Lett.} {\bf A7} (1992) 71;
U. H. Danielsson, Princeton Preprint PUPT-1199 (1992).

\point
D. Gross and I. Klebanov, {\it Nucl. Phys.} {\bf B352} (1991) 671;
A. M. Sengupta and S. Wadia, {\it Int. J. Mod. Phys.} {\bf A6} (1991)
1961;  G. Moore, {\it Nucl. Phys.} {\bf B368} (1992) 557.

\point
G. Moore and R. Plesser, \lq\lq Classical Scattering in 1+1 Dimensional
String Theory", Yale preprint YCTP-P7-92, March 1992.

\point
A. M. Polyakov, {\it Mod. Phys. Lett.} {\bf A6} (1991) 635;
Preprint PUPT (Princeton) -1289 (Lectures given at 1991 Jerusalem
Winter School);
D. Kutasov, \lq\lq Some Properties of (non) Critical Strings",
PUPT-1272, 1991.

\point
E. Witten, {\it Nucl. Phys.} {\bf B373} (1992) 187;
I. Klebanov and A. M. Polyakov, {\it Mod. Phys. Lett} {\bf A6}
(1991) 3273;
N. Sakai and Y. Tanii, {\it Prog. Theor. Phys.} {\bf 86} (1991) 547;
Y. Matsumura, N. Sakai and Y. Tanii, TIT (Tokyo) -HEEP 127, 186 (1992).

\point
I. R. Klebanov, \lq\lq Ward Identities in Two-Dimensional String
Theory", PUPT-1302 (1991);
D. Kutasov, E. Martinec and N. Seiberg, PUPT-1293, RU-31-43.

\point
M. Green, J. Schwarz and E. Witten, ``Superstring Theory", Volume
1, Cambridge University Press, (1987), p.72.

\point
A. Morozov and A. Rosly, {\it Phys. Lett.} {\bf B195} (1987) 554;
J. P. Rodrigues, {\it Phys. Lett.} {\bf B202} (1988) 227;
S. K. Blau et al., {\it Nucl. Phys.} {\bf B301} (1988) 285;
W. de Beer, A. J. van Tonder and J. P. Rodrigues, {\it Phys. Lett.}
{\bf B248} (1990) 67.

\endpage

\end